# Nonlinear Phase Control and Anomalous Phase Matching in Plasmonic Metasurfaces

Euclides Almeida, Guy Shalem, Yehiam Prior

*Department of Chemical Physics, Weizmann Institute of Science, Rehovot 76100, Israel*

Yehiam.prior@weizmann.ac.il

**ABSTRACT**

Metasurfaces, and in particular those containing plasmonic-based metallic elements, constitute a particularly attractive set of materials. By means of modern nanolithographic fabrication techniques, flat, ultrathin optical elements may be constructed. However, in spite of their strong optical nonlinearities, plasmonic metasurfaces have so far been investigated mostly in the linear regime. Here we introduce full nonlinear phase control over plasmonic elements in metasurfaces. We show that for nonlinear interactions in a phase-gradient nonlinear metasurface a new anomalous nonlinear phase matching condition prevails, which is the nonlinear analog of the generalized Snell's law demonstrated for linear metasurfaces. This phase matching condition is very different from the other known phase matching schemes. The subwavelength phase control of optical nonlinearities provides a foundation for the design of flat nonlinear optical elements based on metasurfaces. Our demonstrated flat nonlinear elements (i.e. lenses) act as generators and manipulators of the frequency-converted signal.

**INTRODUCTION**

Metamaterials are a class of artificial materials whose optical properties can be tailored to exhibit phenomena not commonly found in nature, such as negative refraction[1, 2], electromagnetic cloaking[3] or anomalous refraction[4]. These unique optical properties are frequently engineered by single- or multi-layered nanometric objects, often metallic, fabricated on the surface, or within the volume of 'classical' standard materials. Metasurfaces constitute a particularly interesting and attractive sub-set of such materials leading to the possibility of designing and creating, by means of modern nanolithographic fabrication techniques, flat and ultrathin optical elements[5-9]. Within this group of metasurfaces, nano-plasmonic-based metallic elements are the commonly utilized building blocks, and their optical properties are quite well understood and controlled[7, 8] (see also the recent review by Stockman[10]). To date, the majority of work reported on plasmonic-based metasurfaces dealt with their linear optical properties. Nonlinear metasurfaces have been investigated to a much lesser extent, mostly due to the somewhat limited control over the nonlinearities of the individual plasmonic elements in terms of both the phase and amplitude of the specific frequency response.

A particularly fascinating linear result is the generalization of Snell's law upon reflection (transmission) from a plasmonic metasurface, introduced by Yu et al. [5] . In these surfaces an abrupt phase shift is imposed on the impinging light wavefront over distances much smaller than the optical wavelength[6-8] . For metasurfaces with linear phase gradient, anomalous linear refraction and reflection were demonstrated, and it was shown that the familiar laws of refraction and reflection (Snell's laws) must be modified to account for the transverse phase gradient $d\Phi/dx$ as light crosses the interface between two different media. For such configurations, the refracted angle $\theta_2$ is related to the incident angle $\theta_1$ by the generalized Snell's law of refraction:

$$n_2 sin\theta_2 = n_1 sin\theta_1 + \frac{\lambda_0}{2\pi}\frac{d\Phi}{dx} \qquad (1)$$

Where $n_1$ and $n_2$ are the refractive indices of the respective media and $\lambda_0$ is the vacuum wavelength of the light. This generalization of Snell's law is a manifestation of the conservation of photon momentum in the transverse direction. The new term is the additional momentum provided by the interface due to the phase gradient[11] . Based on these principles, Aieta et al. have recently demonstrated achromatic lenses by proper design of the phase elements in metasurfaces[12]. One might expect that such a transverse phase gradient at the interface may also provide an additional momentum that must be included in a nonlinear phase matching scheme.

Metal-based metasurfaces are excellent candidates as ultrathin frequency-converting devices as the field enhancement at plasmonic resonances can enhance nonlinearities by many orders of magnitude. (See a review by Kauranen and Zayats[13]). Second Harmonic Generation (SHG), the lowest order nonlinearity, was studied in plasmonic split ring resonators[14], in nanocups[15], and more recently in nanocavities, individual[16] and coupled[17]. Segal et al.[18] demonstrated control of light propagation including lenses, using a photonic crystal configuration of split ring resonators. Other nonlinear plasmonic responses were measured in the coupling to intersubband transitions[19] in semiconductors and from silver nanoparticles[20] The next, third order nonlinearity was not investigated much either, although, in principle, it exists for any structure and for any material or surface symmetry. Renger et al. [21] studied FWM enhancement by surface gratings, Genevet et al. [22] discussed FWM enhancement by plasmonic gratings, Suckowski et al.[23] demonstrated Four Wave Mixing (FWM) in cross shaped antennas, and discussed phase matching in zero index materials, Zhang et al. [24]

observed enhancement by clusters and discussed Fano resonances in such structures, Maksymov et al. [25] theoretically analyzed FWM in tapered antennas [25] and Simkhovich et al. [26] studied potential super-resolution in plasmonically enhanced FWM . However, none of these works discuss the nonlinear phase response of subwavelength plasmonic structures.

Here we demonstrate full phase control over nonlinear optical interactions in plasmonic metasurfaces. This control is achieved by introducing a spatially varying phase response at the frequency of the nonlinear signal in a metallic metasurface consisting of nonlinear nanoantennas. For such metasurfaces, we derive a new, anomalous nonlinear phase matching condition which differs from conventional phase matching schemes in nonlinear optics. The complete phase control over the nonlinear emission enables us to design of flat nonlinear optical components, such as ultrathin nonlinear lenses with tight focusing, which act as generators and manipulators of the frequency-converted signal.

**RESULTS**

A coherent wave-mixing process obeys the phase matching condition $\Delta \boldsymbol{k} = 0$, where $\Delta \boldsymbol{k}$ is the vectorial sum of the momenta of all photons, incoming and generated, participating in the nonlinear mixing. This phase matching condition determines the direction of the coherent emission, and is particularly useful for spatially identifying and filtering the signal of interest in coherent nonlinear optical spectroscopies. Typically, for efficient wave nixing one uses phase matching schemes such as birefringence, temperature tuning or quasi-phase matching in periodic poled crystals[27, 28]. The latter is attained by designing a nonlinear material with periodic reversal of the sign of the nonlinearity in the propagation direction. In this way, the phase matching condition is $\Delta \boldsymbol{k} + \boldsymbol{G} = 0$, where $\boldsymbol{G}$ is the reciprocal wavevector associated to the longitudinal phase grating.

In FWM, for collinear propagation of the input beams in a homogeneous and nondispersive medium, the phase-matched generated beam propagates in the same direction. This is also the case for the quasi-phase matching scheme, where the poling is in the propagation direction. Generally speaking, and for non-collinear input beams, a more elaborate phase matching scheme is required, and many such schemes are in use[29]. For our metasurfaces, the nonlinear phase gradient imposed by the design of the plasmonic antennas determines the emission phase matched direction.

Optical nanoantennas, like other driven oscillators, reradiate the incoming light at the same frequency[30] but with a shifted phase that changes abruptly across the resonance[5] In this work we use rectangular nanocavities in thin gold films, and make use of the strong intrinsic third-order nonlinearity of the noble metal[13]. To simplify the discussion, and in order to convey the new physical principles more clearly, the Aspect Ratio (AR) was maintained as our single tuning parameter. The total area of the rectangles was kept more or less constant (subject to fabrications imperfections), and the dimensions used in the simulation are the exact values derived from SEM measurements of the actual cavities used in the experiment.

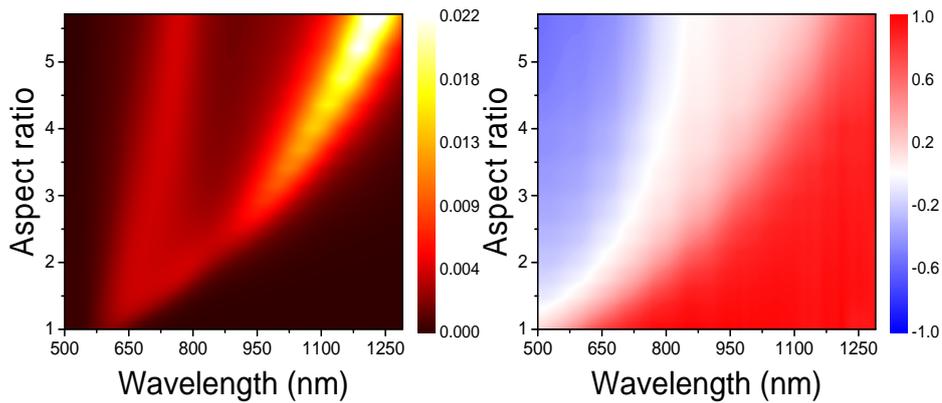

Figure 1: Transmission through rectangular gold nanocavities with varying aspect ratios. a) Linear transmittance spectral intensity and b) corresponding linear spectral phase (color code is in units of $\pi$ )

In Figure 1 we show the calculated linear transmittance spectrum (intensity and phase) for light polarized along the short axis of the rectangles for a set of rectangular nanocavities of different aspect ratios within a free-standing 250-nm thick gold film. In the intensity profile shown in Fig 1a, two distinct cavity resonances are seen: one centered around 650 nm, essentially independent of the AR, which is assigned to a Fabry-Perot (FP) mode[31] and a second one, which is AR-dependent and varies continuously from 650 nm to 1250 nm, and which is attributed to a Localized Surface Plasmon (LSP) excitation[31]. Fig. 1b depicts the phase accumulated for the different wavelengths. The correlation between the LSP mode and the phase shift acquired by the transmitted wave is clear.

Consider now a four-wave mixing configuration where two transform-limited laser pulses $E_1(r,t) = \epsilon_1(r,t) e^{i(k_1 \cdot r - \omega_1 t + \Phi_1)}$ and $E_2(r,t) = \epsilon_2(r,t) e^{i(k_2 \cdot r - \omega_2 t + \Phi_2)}$, travelling at the $k_1$ and $k_2$ directions respectively, interact with a metallic nanoantenna to generate a FWM signal $E_{FWM}(r,t) = \epsilon_{FWM}(r,t) e^{i(k_{FWM} \cdot r - \omega_{FWM} t + \Phi_{FWM})}$ travelling at the $k_{FWM}$ direction and with frequency $\omega_{FWM} = 2\omega_1 - \omega_2$. (See the Methods section for further experimental details) In the frequency domain, the third order polarization (which is the source term for the nonlinear signal at $\omega_{FWM}$) induced at the location $r$ on the nanoantenna is given by[32]:

$$P_3(r,t) = \frac{1}{(2\pi)^3} \int d\omega_1 \int d\omega_1 \int d\omega_2 \chi^{(3)}(\omega_{FWM}, 2\omega_1, -\omega_2) E_1(r, \omega_1) E_1(r, \omega_1) E_2^*(r, \omega_2) \quad (2)$$

Where $\chi^{(3)}$ is the third-order susceptibility of the metal, the fields $E_i(r, \omega_i)$ are position-dependent fields which are affected by the plasmon resonance. Alternatively, we can approximate the nanoantenna as a point dipole and assume that the plasmon resonance leads to effective fields[33], $E_i(\omega_i) = A_i(\omega_i) \epsilon_1(\omega_i) e^{i\Phi(\omega_i)}$ where $A_i(\omega_i)$ is the field enhancement (a real quantity) and $\Phi(\omega_i)$ is the phase response, which was calculated in Figure 1. Therefore, we can rewrite the Equ. 2 as

$$P_3(t) \propto \int d\omega_1 \int d\omega_1 \int d\omega_2 S^{(3)}(\omega_{FWM}, 2\omega_1, -\omega_2) |A_1(\omega_1)|^2 A_2(\omega_2) \epsilon_1(\omega_1) \epsilon_1(\omega_1) \epsilon_2^*(\omega_2) e^{i(2\Phi(\omega_1) - \Phi(\omega_2))} \quad (3)$$

Where $S^{(3)}$ is the effective third-order nonlinear susceptibility of the nanoantenna. The nonlinear FWM signal carries the frequency response at the fundamental frequencies through the phase factor $e^{i(2\Phi(\omega_1) - \Phi(\omega_2))}$. $S^{(3)}$ is a complex quantity, with an imaginary part which is most pronounced when close to the nanoantenna resonance. Furthermore, the field radiated by the nanoantenna will gain an additional phase depending on the geometry of the nanoantenna. The NL phase response of the nanoantennas can be directly calculated using full-wave nonlinear finite-differences time domain (NL-FDTD) calculations.

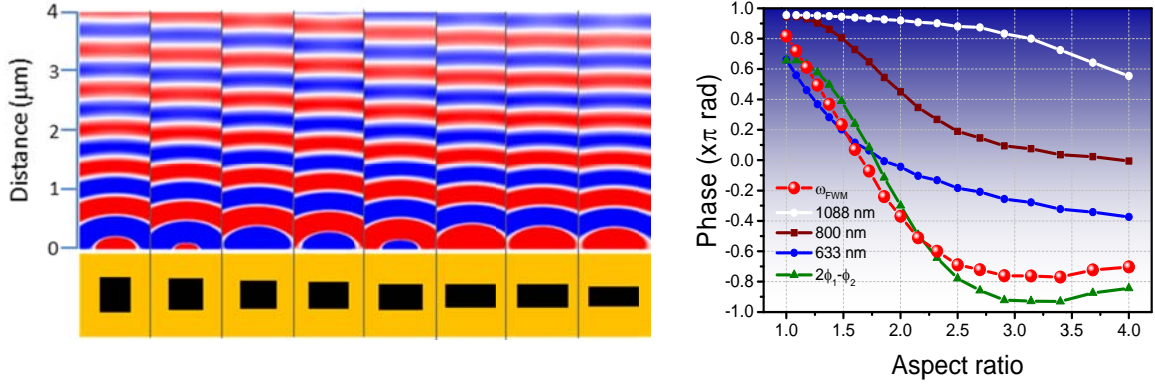

Figure 2: calculated electric field $E_y(\omega_{FWM})/|E_y(\omega_{FWM})|$ for rectangular nanocavities with varying AR. a) The generated FWM at the exit from the film containing the nanocavities as it propagates away from the surface b) The phases accumulated by the different fields (input and generated) upon crossing the metasurface is plotted against the AR: 1088 nm – white circles, 800 nm – brown squares, 633 nm (independently propagating) – blue circles, generated FWM at 633 – red circles, and the phase accumulated at $2\Phi_1 - \Phi_2$ - green triangle (see text)

In figure 2, we show the calculated FWM electric field generated at $\omega_{FWM}$ for a set of nanorectangles with varying AR. In the NL-FDTD calculation, two 60-fs long transformed-limited co-propagating pulses, with center frequencies $\omega_1 = 800\ nm$ and $\omega_2 = 1088\ nm$ are temporally and spatially overlapped at a single nanocavity. At long propagation distances, the wavefront of the FWM signal is a plane wave, and a relative phase-shift for the different nanocavities is extracted. The calculated relative phase shifts for the different waves are shown in figure 2b. The figure depicts the linear phase shift accumulated by individually propagating waves at 800nm (brown), 1088nm (white) and 633nm (blue), and then the nonlinear phase accumulated by the generated FWM beam at 633nm (red). The calculated $\Phi_{FWM}$ does not fit the phase of the 633nm wave; it is better represented by the phase difference $2\Phi_1 - \Phi_2$. The calculated phase difference provides a fairly good estimate for the phase of the generated FWM signal, but the fit is not perfect. For a better description of the generated phase two additional factors need be included. The first, and less critical one, is integration over the laser bandwidth, which for these ultrashort pulses is not negligible. The more important factor is the fact that the FWM signal is generated gradually over the length of the nanocavity, so that the phase accumulated directly by this wave, as it is building up, should also be effectively included. The linear phase response of the cavity at the FWM frequency may be numerically calculated in detail, or can be included as an effective dielectric constant $\epsilon_{eff}$ that appears in the nonlinear wave equation

$$\nabla^2 \boldsymbol{E}_{FWM} - \frac{\epsilon_{eff}(\omega_{FWM})}{c^2}\frac{\partial^2}{\partial t^2}\boldsymbol{E}_{FWM} = \frac{1}{\varepsilon_0 c^2}\frac{\partial^2}{\partial t^2}\boldsymbol{P}^{(3)} \qquad (4)$$

As mentioned, our direct NL-FDTD calculation provides the phase $\Phi_{FWM}$, and therefore we will use results such as of Fig. 2b for the design of our metasurface.

In analogy to the linear phase gradient arrays [5] we design an array with a phase gradient at the frequency of the nonlinear signal, and with this structure we demonstrate the new phase matching condition for gradient metasurfaces. For comparison, we also measure a 'uniform' array, where all nanocavities have the same Aspect Ratio.

Consider the FWM configuration illustrated in Fig. 3. The same two ultrashort pulses, with wavevectors $\boldsymbol{k}_1$ and $\boldsymbol{k}_2$ respectively, are now spatially and temporally overlapped and focused on the graded-phase metasurface to generate a FWM signal at $\omega_{FWM} = 633\ nm$ and $\boldsymbol{k}_{FWM}$. After the sample, the fundamental beams are filtered and the FWM signal is imaged on a CCD camera which records its k-space information. We begin with a measurement of the generated FWM from two different metasurfaces – each consisting of four rectangles 450 nm apart, in the uniform case all with AR=1.9, and for the phase gradient one the aspect ratio covers the range of AR= 1.1 - 2.9 .

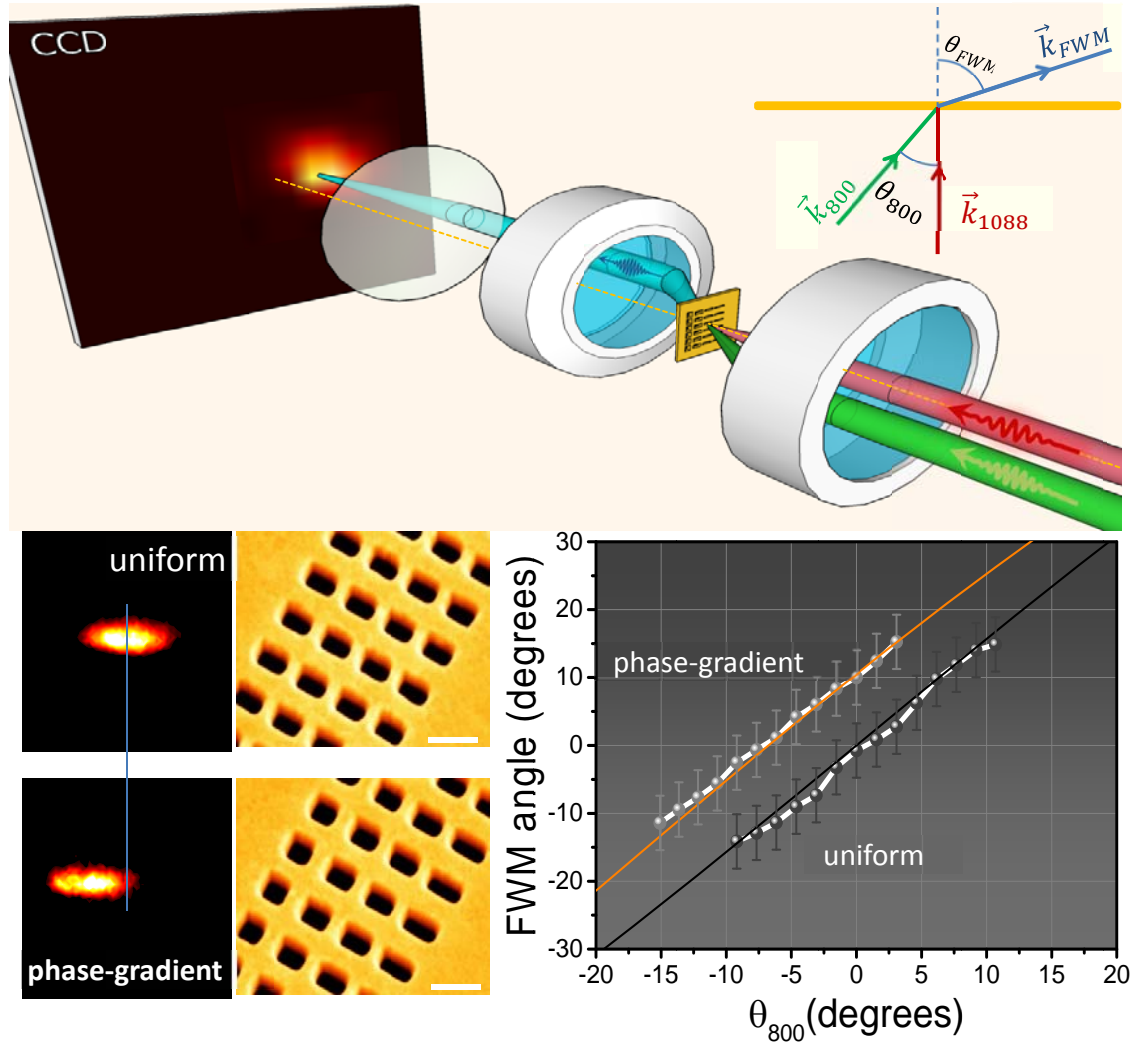

Figure 3: k-space analysis of the FWM. a) Optical arrangement for measuring the FWM angle dependence. The position of the 800nm (green) beam on the focusing lens, as determined by a translation stage, controls the input angle. The θ$_{FWM}$ was determined in relation to the beam generated by a uniform structure. b) CCD image for a signal from a uniform unit cell, and c) from a phase gradient unit cell  d) Input angle dependence of the phase matching angle for the uniform and phase-gradient metasurfaces. The orange line is the line fit to the anomalous phase matching condition (Equ. 6), while the black line depicts the conventional phase-matching condition.

We measure the FWM output angle as a function of the input angle $\theta_{800}$ of the $\omega_1 = 800\ nm$, beam for both the uniform and phase gradient metasurfaces. The incident angle is controlled by a parallel and lateral displacement of the position of the beam on the back lens of the focusing objective, while the angle of incidence of the $\omega_2$ beam is kept at normal incidence. Figure 3 depicts the results for both metasurfaces: for the phase gradient surface, the FWM signal is emitted at a different angle which is ~10 degrees higher than the signal from the uniform surface, indicating a different phase matching condition. This new phase matching condition includes the additional momentum provided by the metasurface, along the phase gradient direction:

$$\bm{k}_{FWM}^{new} = \bm{k}_{FWM} + \Delta \bm{k}_x \qquad (5)$$

The net momentum provided by the metasurface to the FWM signal is related to the phase gradient by $\Delta \bm{k}_x = (d\Phi_{FWM}/dx)\bm{u}_x$, and includes the momentum given by the metasurface to all the beams participating in the nonlinear conversion process.

This new, anomalous phase-matching condition for FWM assumes the form:

$$\bm{k}_{FWM}^{new} = \bm{k}_{FWM} + \frac{d\Phi_{FWM}}{dx}\bm{u}_x \qquad (6)$$

Equ. 6, combined with the NL-FDTD calculation for $\Phi_{FWM}$ provides a framework for the design of phase-gradient nonlinear metasurfaces. In figure 3, the orange curve represents a nonlinear fit to Equ. 6, from which we extract a value for the phase gradient provided by the metasurface over a unit cell, in this case $\Delta\Phi_{FWM} = -0.55\pi$.

With the proper choice of $d\Phi_{FWM}/dx$, one can control the beam steering of the FWM emission. In figure 4 we show a NL-FDTD calculation, using parameters similar to the experimental, of the angle dependence of the FWM signal for a series of phase-gradient metasurfaces (circles). In the calculation, the field close to the exit surface is projected to the far-field so that the FWM angle can be extracted. The line fits to Equ. 6 (blue lines) are also plotted and are in good agreement to the NL-FDTD calculations.

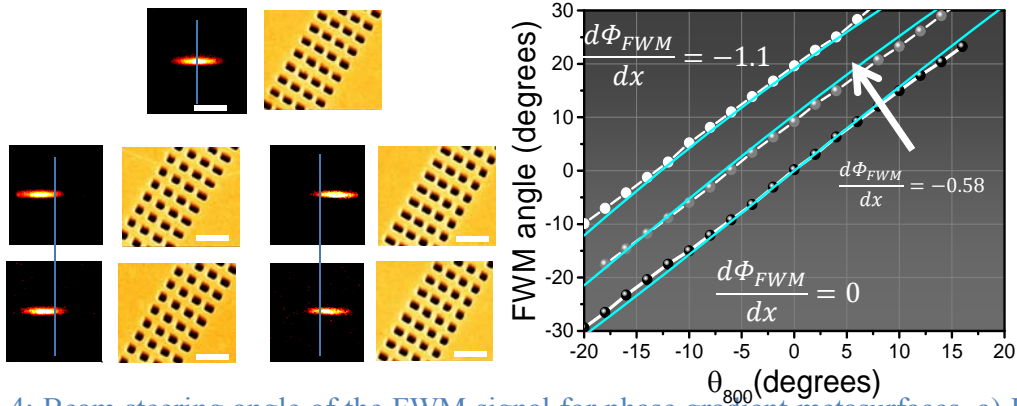

Figure 4: Beam steering angle of the FWM signal for phase gradient metasurfaces. a) Individual phase gradient structures and the generated FWM at normal incidence. Top – uniform structures; Left – AR increasing to the right' Right – AR increasing to the left b) Measured and calculated FWM angle dependence for phase gradient structures, where the phase gradient was taken from figure 2 and the calculated blue lines are derived from the anomalous phase-matching condition (Equ. 6).

The observations described so far were done with a single unit cell over which the phase changes. Naturally, if several (many) unit cells are arranged in a periodic manner, they form a blazed grating. If the phase change across a unit cell in a periodic array is different from $2\pi$, the analysis of the anomalous phase-matching condition must be modified to include additional effects such as the general theory of diffraction[34] in the linear case, or the Raman-Nath diffraction[35, 36] in the nonlinear one. Thus was demonstrated for SHG from gratings with a periodic modulation in the sign of the nonlinear susceptibility in the transverse direction [18, 37]. The gratings discussed here are different – they behave like blazed mechanical or holographic gratings which are not symmetric in terms of 'positive and 'negative' transverse directions. For such gratings, the positive and negative diffraction orders are different, and the negatively diffracted orders (m= -1, -2 …) exhibit much lower diffraction intensity. To illustrate this more general situation, the k-space analysis of the FWM emission from an array of periodic unit cells is shown in figure 5. The spots seen on the CCD images for collinear, normal excitation are explained by the different orders of diffraction of the blazed grating. The angle of diffraction of the different diffraction orders is determined by the grating period, and agrees with the Raman-Nath diffraction formula $sin\theta_m = m\lambda_{FWM}/\Lambda$. High order diffraction modes are also seen, but with weaker intensity compared to the zeroth and first modes, also in accordance with the Raman-Nath diffraction theory. Experimentally, the NA of the collection optics only allows a limited field of view, and therefore here we show only the first order.

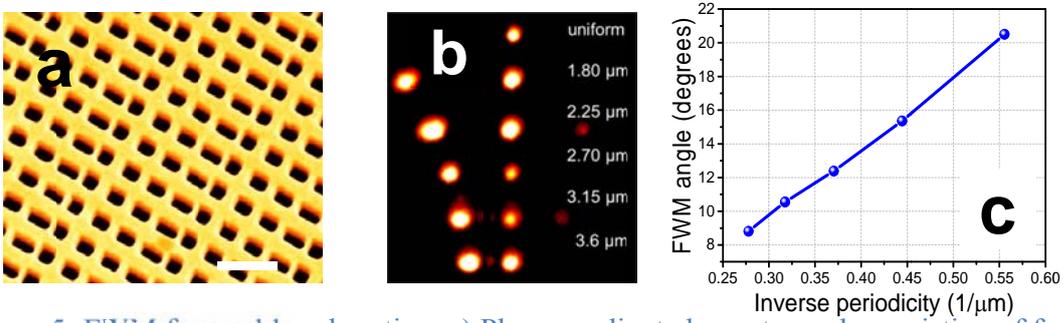

Figure 5: FWM from a blazed grating. a) Phase gradient elements, each consisting of four elements with AR= 1.1 , 1.5 , 1.9 , 2.9   b) CCD images for the zeroth and first diffraction orders for different periodicities d) Measured FWM emission angle as a function of the grating periodicity.

To illustrate the power and flexibility of these nonlinear phase gradient metasurfaces we designed a nonlinear meta-lens which focuses the wavelength of choice in a specific FWM configuration. The ultrathin lens operates by imposing a radially-dependent relative phase shift on the wavefront [4]:

$$\Phi(r) = \frac{2\pi}{\lambda_0}\left(\sqrt{r^2 + f^2}\right) \quad (7)$$

Here $f$ is the desired focal distance of the lens and $\lambda_0$ is the free-space wavelength. The individual elements of the nonlinear nanoantennas must be located and designed (choice of AR) with relative FWM phases $\varphi_{FWM}$ in accordance with the above equation. Our lenses were designed and fabricated on 250 nm thick free-standing gold films, with the design focal length ranging from 5-30 μm. The lens was designed for operation at $\omega_{FWM} = 2\omega_1 - \omega_2 = 633\ nm$, with $\omega_1 = 800\ nm$ and $\omega_2 = 1088\ nm$. The relative phase required for each individual element is shown in Figure 6a, along with the corresponding aspect ratios of the nanoantennas. The metalens is made of 11 concentric ring of the rectangular nanoantennas. The experimentally observed and numerically calculated tomographic images of the focal regions are shown in figure 6. The FWM signal is focused to a Gaussian spot with full-width at half maximum FWHM = 1.8 μm.

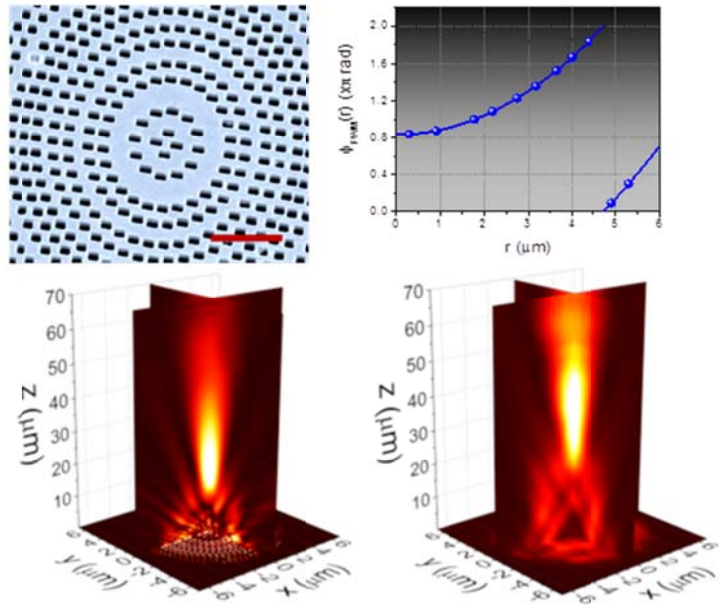

Figure 6: A nonlinear metalens of focal distance f = 30 μm based on FWM operating at ω_FWM=633 nm. a) SEM image of the fabricated ultrathin metalens. The scale bar is 2 μm. b) Relative nonlinear phase c) Experimental (right) and NL-3D-FDTD simulated (left) images of the focal region.

Other nonlinear metalenses of different focal distances were also designed and fabricated and are shown in figure 7 for $f$ = 5, 10 and 60 μm respectively. All lenses provide tight focusing, with FWHM = 1.0, 1.0 and 2.2 μm for $f$ = 5, 10 and 60 μm respectively.

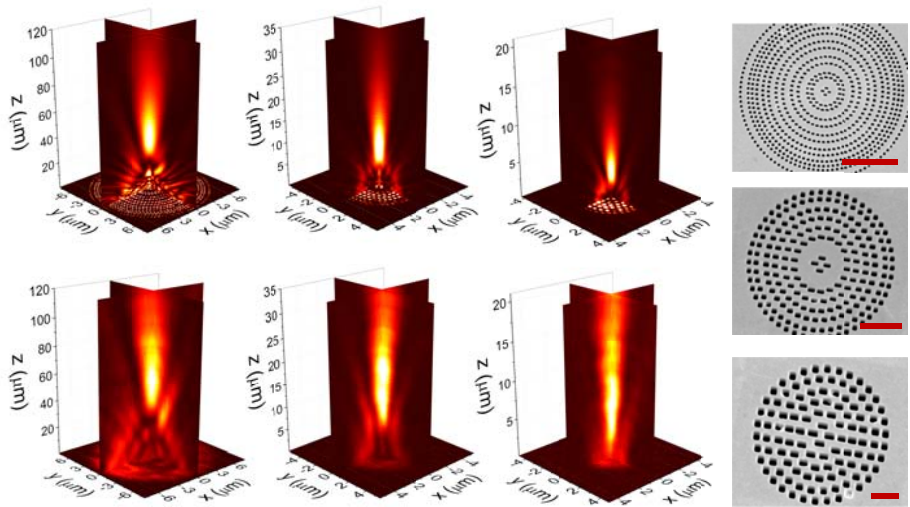

Figure 7: Nonlinear ultrathin plasmonic metalenses, designed for different focal distances (left to right) f = 60, 10, 5 μm. 3D tomographic images of the simulated (top) and experimentally measured (bottom) focal region light. The actual lenses are shown on the right, where the scale bar is 5, 2 and 1 microns for f=60, 10 and 5 respectively.

**DISCUSSION**

In the present work we demonstrate, for the first time, control over the nonlinear phase and beam steering in phase gradient metasurfaces. As in the linear regime, where the laws of refraction and reflection had to be modified, the phase control of nonlinear nanoantennas leads to modifications in the laws governing nonlinear phenomena, such as NL scattering, NL refraction and frequency conversion. Several ultra-thin linear optical components[7, 8], such as lenses[38, 39], holograms[40] and broadband waveplates[41] using gradient metasurfaces have been proposed and demonstrated and they are all based on the abrupt phase changes experienced by light upon propagation through a properly designed ultrathin layer of metamaterial. We have carried this concept one step further, designed and fabricated blazed phase gradient metasurfaces, and carefully analyzed their nonlinear optical properties. We show that FWM from such metasurfaces reveals several new phase properties: scattering from a phase gradient unit cell which depends on the phase accumulated over the unit cell; scattering from periodic structures, where each individual constituent is a phase gradient unit cell; New, anomalous phase matching condition for FWM from surfaces with built in phase gradient.

In addition we designed and fabricated ultrathin FWM, lenses, and demonstrated tight focusing with focal lengths of several microns. These nonlinear metalenses provide shorter and tighter focusing than nonlinear metalenses based on SHG [12]. Furthermore, the lenses which operate through the FWM process, which depends on the third-order nonlinearity of the material, do not have any restrictions on the symmetry of the design which is characteristic to elements based on SHG. These nonlinear metalenses can be integrated in light detectors based on frequency conversion to provide more sensitive detection.

In all these cases, the experimental results were compared to numerical solutions (FDTD, Lumerical solutions package[42]) of the wave equations with nonlinear terms added to them. The agreement is generally very good, and whenever relevant, comparison to analytical expressions is added.

The intensity of the NL signal generated in each individual rectangular nanocavity used in this work depends on the aspect ratio of the rectangle, which may lead to distortions to the wavefront of the NL signal. This drawback may be circumvented by changing other parameters such as area of the hole, or shapes presenting multiple resonances such as V-shaped antennas [5]. Furthermore, nonlinear metasurfaces may have potential applications in high sensitivity nonlinear sensing, such as surface-enhanced CARS[43].

In conclusion, we have demonstrated that nonlinear phase control of metasurfaces leads to a new phase-matching condition for frequency conversion on nonlinear optics. Using NL-FDTD calculations, we show that rectangular metallic nanocavities can act as nonlinear nanoantennas that are capable of providing a continuous phase gradient at the frequency of the nonlinear signal. Using k-space analysis, we were able to measure deviations of the wavevectors of the nonlinear signals from the values expected in conventional phase matching. We modified the phase matched condition to include the phase-gradient, which was interpreted as the additional transverse momentum given by the metasurface. We discussed the differences between this approach and the Raman-Nath theory of nonlinear diffraction. Additionally, we demonstrated an application of the phase control of nonlinear nanoantennas by designing and fabricating nonlinear metalenses based on FWM, which can generate and manipulate the wavefront of the NL signal to focus tightly the frequency converted radiation. The phase control of nonlinear nanoantennas may have technological implications for the next generation of efficient nonlinear metamaterials with complex functionalities.

**METHODS**

**Sample fabrication**

The samples were fabricated by FIB milling on a high quality free-standing gold film. The procedure for fabrication of free-standing gold films is described in details in [31] Briefly, using an e-beam evaporator, we deposited a 10 nm thick Cr adhesion layer and a 250 nm thick gold layer on a polished silicon wafer. On the opposite side of the wafer, we had previously grown a circular $Si_3N_4$ mask by plasma-enhanced chemical vapor deposition. The mask was chemically etched using KOH and the remaining free-standing metallic area was etched with HCl to remove the adhesion layer.

**Linear FDTD simulations**

The transmittance spectrum and the relative spectral phase response of the rectangular metallic nanocavities were calculated using the commercial software Lumerical FDTD solutions. The values of the dielectric constants were taken from the data table of Gold from Palik[44].

**Nonlinear FDTD simulations**

The nonlinear phase was calculated using the nonlinear material implementation of Lumerical. The base material is Palik gold which is assumed to have instantaneous (non-dispersive) third-order nonlinearity $\chi^{(3)} = 10^{-18}$ m$^2$/V$^2$. As input light sources we used two temporally overlapped transform-limited plane wave sources centered at $\omega_1 = 800\ nm$ and $\omega_2 = 1088\ nm$, with pulse duration 60 fs, propagating parallel to the z-direction. The polarization of both pulses is perpendicular to the long axis of the rectangles. The y-component of the real and imaginary parts of the electric field (i.e. the propagating waves) of the FWM signal is recorded on a y-normal plane spanning the whole simulation area. The dimensions of the mesh were set to dx=dy=dz=5 nm and perfectly matched layers were added in all dimensions.

**K-space analysis**

For the k-space analysis of the FWM signal from phase-gradient antennas, the same simulation parameters used in the phase response were kept, except now the $\omega_1 = 800\ nm$ source propagates in the x-z plane with a variable incidence angle $\theta_{800}$ with respect to the normal. The exit fields on the opposite side of the metasurface are recorded in a z-normal plane and projected to the far-field, where the angle $\theta_{FWM}$ of the FWM signal at $\omega_{FWM}$= 633 nm is calculated as a function of $\theta_{800}$.

**Nonlinear lenses**

For the design of our lenses we kept the same parameters used in the calculations of the NL phase response. However, to decrease the simulation time, we used symmetric boundary condition in the x-dimension and anti-symmetric in the y-dimension. The dimensions of the fine mesh around the lenses were set to dx=dy=10nm and dz=5 nm for the $f$ = 5 and 10 μm lenses and dx=dy=15nm and dz=5 nm for the $f$ = 30 and 60 μm lenses.

**K-space measurements**

In the FWM experiments, we used the setup described in details in[31] An Optical Parametric Amplifier (OPA), pumped by a 1 KHz amplified Ti:Sapphire laser, was used as the light source for the $\omega_2$=1088 nm pulses, while the pulses of the Ti:Sapphire laser that pumped the OPA were used as the fundamental $\omega_1$=800 nm beam. Both $\omega_1$ and $\omega_2$ pulses have the same pulse duration of 60 fs. The beams travel two distinct optical paths, where the intensity and polarization of each individual beam could be controlled by a set of half-wave plate and polarizer in order to avoid optical damage to the samples. Both beams are focused and overlapped in the sample, by an objective lens of numerical aperture N.A = 0.42 (Mitutoyo M Plan Apo 50X Infinity-Corrected). The incident angle $\theta_{800}$ of the $\omega_1$ beam can be varied by controlling the lateral displacement of the beam with computer-controlled translation stage supporting a beam splitter whose primary role was to merge the optical path of two beams. The $\omega_2$ beam at 1088 nm was always kept normal to the surface. The temporal overlap between the two beams as the input angle was varied was monitored and controlled by properly delaying the $\omega_1$ beam. The FWM signal centered at $\omega_{FWM}$= 633 nm is collected by an objective lens with N.A = 0.42 (Mitutoyo M Plan Apo SL50X infinity-corrected) and focused by spherical lens of $f$ = 75 mm onto an EMCCD camera (Andor iXon DV885) The fundamental beams are filtered by a pair of shortpass filters (Thorlabs FES0700 and FESH0750). In the angular measurements in figures 4 and 5, the focusing objective with NA = 0.42 was replaced by a lens of focal length f = 50 mm in order to illuminate a larger area.

**Nonlinear lenses measurements**

The experimental setup for measurements of the nonlinear lenses is a modification of the k-space setup described above. Both $\omega_1$ and $\omega_2$ beams are normally incident. The focusing objective is replaced by a spherical lens with $f$ = 100 mm. The FWM signal out of the lenses is collected by the imaging (NA=0.42) objective and is imaged directly onto the EMCCD in real (physical) space. The imaging objective is supported on a computer-controlled translation stage that can vary the focal plane of the objective and record 3D tomographic images of the focal region.

**ADDITIONAL INFORMATION**

The author(s) declare no competing financial interests.

**ACKNOWLEDGEMENT**

This work was funded, in part, by the Israel Science Foundation, by the ICORE program, by an FTA grant from the Israel National Nano Initiative, and by a grant from the Leona M. and Harry B. Helmsley Charitable Trust. Discussions with Roy Kaner, Yaara Bondy and Yael Blechman are gratefully acknowledged.

**AUTHORS CONTRIBUTION STATEMENT**

All authors conceived the idea. EA fabricated the samples and performed most of the experimental work. EA and GS performed the numerical simulations. All authors jointly wrote the paper and contributed to the physical understanding of the phenomena described.